\documentclass[11pt]{article}
\usepackage{jheppub}
\usepackage[utf8]{inputenc}
\usepackage{amsmath}
\usepackage{graphicx}
\usepackage{subcaption}
\usepackage{xcolor}
\usepackage{comment}
\usepackage[toc,page]{appendix}
\usepackage{feynmp}
\usepackage{tikz-feynman}
\tikzfeynmanset{compat=1.1.0}
\usetikzlibrary{positioning,arrows}
\usetikzlibrary{decorations.pathmorphing}
\usetikzlibrary{decorations.markings}

\graphicspath{{Images/}}
\newcommand{\ak}[1]{{\color{black} #1}}

\title{\ak{Universality of} Dissipation across \ak{Holographic} Interfaces}
\date{\today}
\author{Andreas Karch${\rm ,}^{1}$ and}
\author{Mianqi Wang${}^{1}$}

\affiliation{${}^{1}$Theory Group, Weinberg Institute, Department of Physics,
University of Texas \\  2515 Speedway, Austin, TX 78712, USA.}

\emailAdd{karcha@utexas.edu}
\emailAdd{mqwang@utexas.edu}

\abstract
{Motivated by recent results in spin chains we study dissipation and relaxation in a two-dimensional \ak{holographic} interface conformal field theory (ICFT) in which degrees of freedom on one side of the interface are coupled to an external bath, while the other side remains isolated. In the bulk description this setup is realized by gluing a supersymmetric Janus geometry to a BTZ black hole region, with the coupling implemented through a double-trace deformation. We determine the  quasinormal modes in the bulk by solving the double-trace matching conditions of the system and bath. Focusing on the fixed AdS$_2$ descendant channel at large frequency, we introduce the dimensionless ratio $c_{UV}$
as a projected measure of interface-induced suppression of dissipation rate. Analytically we find that in this limit, $c_{UV}$ is independent of coupling details to the bath. It is a strong candidate for a universal interface observable characterizing dissipation and relaxation across the interface, in additional to the $c_{\rm relax}$ raised in earlier work for lattice models.}

\begin{document}
\tikzset{
line/.style={thick, decorate, draw=black,}
 }
\maketitle

\section{Introduction}
Conformal interfaces in two-dimensional quantum field theories provide a rich arena for probing universal properties of strongly coupled systems. In two-dimensional interface conformal field theories (ICFTs), an interface separates two conformal fixed points while preserving a diagonal conformal symmetry. Despite their local nature, such interfaces are characterized by a small set of universal quantities that capture intrinsic properties independent of microscopic details. Well-known examples include the effective central charge $c_\text{eff}$, the energy transmission coefficient $c_{LR}$, and the boundary entropy or g-factor, all of which admit precise definitions within field theory and have elegant realizations in holographic duals \cite{PhysRevLett.67.161,Quella_2007,Meineri_2020,Karch:2022vot,Karch:2023evr,Chiodaroli:2010ur,Chiodaroli:2016jod}.

From the bulk perspective, these universal ICFT data are naturally \ak{realized} in Janus geometries, where the bulk scalar profile interpolates between distinct asymptotic AdS regions. In this framework, the interface observables can be extracted geometrically from minimal surfaces calculating  entanglement entropies, or from a bulk mode analysis.  Supersymmetric extensions, including super-Janus solutions, further enrich this picture by allowing controlled embeddings into string theory and enabling the study of supersymmetric ICFTs with exact control over protected quantities.

While much of the existing literature has focused on static or equilibrium characteristics of conformal interfaces, considerably less is known about their dynamical and dissipative properties. In particular, when an interface system is coupled to an external environment, new universal structures may emerge that are not visible in equilibrium observables. Understanding such dissipative dynamics is of both conceptual and practical interest, as it connects interface CFTs to broader questions of open quantum systems, relaxation, and information loss at strong coupling.

Recently, progress in this direction was made in \cite{barad2025dissipationmeetsconformalinterface}, where dissipation at conformal interfaces was studied using free fermions on the lattice both numerically and perturbatively. A central insight of that work is that the late-time relaxation of interface degrees of freedom can be characterized by a universal dissipation rate governed by a Liouvillian gap \cite{Baumgartner_2008}. This suggests the existence of new interface data, \ak{$c_\text{relax}$}, controlling nonequilibrium dynamics, analogous in spirit to $c_\text{eff}$, $c_{LR}$, \ak{and} $g$, but intrinsically dynamical in nature.

Concretely, the authors of \cite{barad2025dissipationmeetsconformalinterface} use the Lindblad framework to describe the coupling to the bath. Their jump operator is localized at one site of the fermion chain. The jump operator gives rise to non-unitary time evolution, meaning that states do not simply oscillate but excitations exponentially decay to a steady state. Since the system is regulated both in the UV, by being put on the lattice, and in the IR, by being studied at finite volume, there is a finite set of eigenmodes and one can identify a Liouvillian gap from the smallest decay rate. While this quantity itself depends strongly on the details of the jump operator, most notably its coupling strength, the ratio of the gap between the system with and without interface is universal. This ratio is defined as $c_\text{relax}/c$. Naturally the presence of the interface shields the side without bath coupling from the dissipation and correspondingly $c_\text{relax}/c$ turns out to be less than one. In the case that the interface separates the system into two decoupled boundary CFTs $c_\text{relax}$ vanishes as the half of the system without dissipation has only unitary evolution and hence zero decay rate.

Motivated by these developments, we couple the BPS-Janus supergravity
background \cite{Chiodaroli_2010,Chiodaroli:2016jod} to a thermal BTZ bath by a
uniform double-trace deformation on one side of the interface.  The projected
damping parameters studied here decrease as the radial excitation energy is
increased, so they do not strictly define a nonzero Liouvillian gap.  We therefore
define $c_{\text{UV}}$ more narrowly, as the ratio of the projected decay rates
of fixed-$k$ radial modes at matched large physical radial energy, with and
without the Janus deformation.  The denominator is the corresponding projected channel in the
$\kappa=1$ reference geometry, with the same renormalized boundary condition.
This definition is not a minimum over the complete coupled spectrum.

Within this projected radial-UV channel we find
$c_{\text{UV}}/c=1/\kappa$, equal to $c_{\mathrm{eff}}/c$ for this half-BPS
super-Janus family.  The limiting ratio is independent of the
double-trace coupling, the bath temperature, and finite subtraction constants,
although the individual projected damping parameters depend on these choices
and vanish as
$1/\log^2 n$.  This channel result is suggestive of dynamical interface data,
but it does not establish that the projected root is a quasinormal-mode pole of
the full infinite matrix when descendant labels scale with frequency.  The
relation to the regulated free-fermion quantity $c_{\text{relax}}$ therefore
remains a question about the complete spectrum.

This paper is organized as follows.  In Sections \ref{sec:ads3}, \ref{sec:bps},
and \ref{sec:btz} we review the modes of pure AdS$_3$, BPS-Janus, and BTZ.  In
Section \ref{sec:couple} we formulate the exact half-space double-trace problem.
In Section \ref{sec:cuv} we derive the fixed-$k$ projected radial-UV limit and
define $c_{\text{UV}}$.  Section \ref{sec:future} discusses future directions.

\paragraph{Note added.}
An earlier version of this paper pursued a different approach, based on a numerical analysis of the full quasinormal mode spectrum and the extraction of a relaxation quantity $c_{\mathrm{relax}}$ from its least-damped pole. The present version instead develops an analytic treatment of the large-$n$ limit with fixed AdS$_2$ descendant number $k$, leading to the projected ultraviolet quantity $c_{\mathrm{UV}}$. Because the corresponding damping rates vanish in this limit, the present analysis does not by itself determine a finite Liouvillian gap or establish the relation between $c_{\mathrm{UV}}$ and $c_{\mathrm{relax}}$, which are left to future work.

\section{(Quasi)-normal modes for the various geometries}
\label{sec:normal-modes}

Before understanding how the coupling of system and bath affects the spectrum of fluctuations, we first need to collect what is known about the fluctuations in the various geometries we employ before we couple them to each other.

\subsection{Normal modes of empty AdS$_3$}
\label{sec:ads3}
In this subsection, we review the normal modes of empty AdS$_3$ without an interface. Although the results are standard, in order to map the results to Janus geometry, we approach it in \ak{AdS$_2$} slicing coordinates rather than the usual cylinder picture.

Consider the foliation of asymptotic \ak{AdS$_3$ by AdS$_2$ slices}
\begin{equation}
    ds^2=dx^2+e^{2A(x)}\,\sec^2\sigma\,( -dt^2+d\sigma^2)
\end{equation}
$x\to\pm\infty$ are boundary pieces from left and right, and $\sigma\in[-\tfrac{\pi}{2},\tfrac{\pi}{2}]$. For empty AdS$_3$ we have $e^A=\cosh x$. Throughout the paper \ak{the} AdS$_3$ radius is \ak{set to} $\ell_{AdS}=1$. A probe scalar with mass $M$ in AdS$_3$ solved by the Klein--Gordon (KG) equation can be decomposed as 
\begin{equation}
    \Psi_\omega = \sum_n f_n(x)g_{\omega,n}(\sigma)e^{-i\omega t}
    \label{eq:adssol}
\end{equation}
where $g_{\omega,n}$ are the eigenfunctions of the AdS$_2$ radial KG equation with mass $m_n$, and $f_n(x)$ solves the slicing modes equation. Following results of \cite{Aharony_2003,Karch_2017}, for general warpfactor $e^{A(x)}$ the eigenvalue equation for the slicing direction is
\begin{equation}
    f_n''(x)+2A'(x)f_n'(x)=(M^2-e^{-2A(x)}m_n^2)f_n(x)
    \label{eq:radialode}
\end{equation}
For empty AdS$_3$, the mass tower is
\begin{equation}
    \Delta_n = \Delta+n,\quad n=0,1,\dots
\end{equation}

where $m_n^2 = \Delta_n(\Delta_n-1)\equiv\nu_n(\nu_n+1)$ and $\Delta\equiv\Delta_+=1+\sqrt{M^2+1}$. The solutions $f$ and $g$ can be written out explicitly \cite{Karch_2017,Higuchi_2022}
\begin{equation}
    \begin{split}        &f_n(x)=\frac{C_n\Gamma(\Delta)}{2\cosh x}P_{-\Delta_n}^{1-\Delta}(\tanh x),\quad C_n^2=\frac{2\,\Gamma(\Delta_n+\Delta-1)(2\Delta_n-1)}{n!\,\Gamma(\Delta)^2}\\
        &g_{\omega,n}^{(1)}(\sigma)=(\cos\sigma)^{\Delta_n}~_2F_1\left(\frac{\Delta_n+\omega}{2},\frac{\Delta_n-\omega}{2};\frac{1}{2};\sin^2\sigma\right)\\
        &g_{\omega,n}^{(2)}(\sigma)=\sin\sigma\,(\cos\sigma)^{\Delta_n}~_2F_1\left(\frac{1+\Delta_n+\omega}{2},\frac{1+\Delta_n-\omega}{2};\frac{3}{2};\sin^2\sigma\right)\\
    \end{split}
    \label{eq:fg}
\end{equation}

The two independent solutions have definite parity under
$\sigma\to-\sigma$, which exchanges the two AdS$_2$ boundary endpoints:
$g_{\omega,n}^{(1)}$ is even and $g_{\omega,n}^{(2)}$ is odd.  Consequently,
their asymptotic coefficients at $\sigma=\pm\pi/2$ are equal for
$g_{\omega,n}^{(1)}$ and opposite for $g_{\omega,n}^{(2)}$.

Near the boundary of AdS$_3$, the normalizable vev $\beta$ and non-normalizable source $\alpha$  defined by
\begin{equation}
    \Psi_\omega\sim \alpha_\omega(\sigma)\, e^{-(d-\Delta)|x|}+\beta_\omega(\sigma)\,e^{-\Delta |x|} 
\end{equation}
are read off by the AdS$_2$ modes near the expansion, via the BOPE formula \cite{Aharony_2003}
\begin{equation}
    \mathcal{O}_{d=2}(\sigma)=\sum_n\frac{b_{\Delta_\mathcal{O}\Delta_n}}{\tan(\tfrac{\sigma}{2}+\tfrac{\pi}{4})^{\Delta_\mathcal{O}-\Delta_n}}\mathcal{O}_{d-1,n}
    \label{eq:bope}
\end{equation}

We derive the AdS$_3$ normal modes by selecting the normalizable branch at both
AdS$_2$ boundary endpoints.  Transforming the hypergeometric function gives
\cite{Higuchi_2022}
\begin{equation}
\begin{split}
        g_{\omega,n}^{(1)}(\sigma) 
= (\cos \sigma)^{\Delta_n} 
\,A_1(\omega)\, F\!\left(
\frac{\Delta_n+\omega}{2}, \,
\frac{\Delta_n-\omega}{2}; \,
\frac{1}{2}+\Delta_n; \,
\cos^{2}\sigma
\right)\\
+ (\cos \sigma)^{1-\Delta_n} 
\, B_1(\omega) \,
F\!\left(
\frac{1-\Delta_n+\omega}{2}, \,
\frac{1-\Delta_n-\omega}{2}; \,
\frac{3}{2}-\Delta_n; \,
\cos^{2}\sigma
\right).
\end{split}
\end{equation}

where 
\begin{equation}
A_1(\omega) :=
\frac{\Gamma\!\left(\tfrac{1}{2}\right)\,
      \Gamma\!\left(\tfrac{1}{2}-\Delta_{n}\right)}
     {\Gamma\!\left(\tfrac{1-\Delta_{n}+\omega}{2}\right)\,
      \Gamma\!\left(\tfrac{1-\Delta_{n}-\omega}{2}\right)},
\qquad
B_1(\omega) :=
\frac{\Gamma\!\left(\tfrac{1}{2}\right)\,
      \Gamma\!\left(\Delta_{n}-\tfrac{1}{2}\right)}
     {\Gamma\!\left(\tfrac{\Delta_{n}+\omega}{2}\right)\,
      \Gamma\!\left(\tfrac{\Delta_{n}-\omega}{2}\right)} .
\end{equation}

\ak{We} perform the same transformation for $g^{(2)}$, giving $A_2,B_2$.
Normalizability requires $B_{1,2}(\omega)=0$.  The even sector gives
$\omega=\Delta_n+2p$, while the odd sector gives
$\omega=\Delta_n+2p+1$, with $p\in\mathbb N_{\geq0}$.  Combining the two
parity sectors gives the discrete empty AdS$_3$ normal modes
\begin{equation}    \omega_{k,n}=\Delta_n+k=\Delta+n+k,\quad k,n\in\mathbb{N}_{\ge 0}
\end{equation}

This is the standard global AdS$_3$ spectrum in cylinder coordinates.  At these
discrete values $\omega=\omega_{k,n}$, the Legendre functions in $g$ become
Jacobi polynomials that form an orthonormal basis (here with
$\nu=\nu_n=\Delta_n-1$)
\begin{equation}
g_{\nu,k}(\sigma) 
= N_{\nu,k} \, (\cos \sigma)^{\nu+1} \,
P^{(\nu+\tfrac{1}{2}, \, \nu+\tfrac{1}{2})}_{k}(\sin \sigma),
\label{eq:ads2}
\end{equation}

where
\begin{equation}
P^{(a,b)}_{m}(x) :=
\frac{\Gamma(m+a+1)}{m!\,\Gamma(a+1)} \,
{}_2F_{1}\!\left(-m,\, m+a+b+1;\, a+1;\, \tfrac{1-x}{2}\right).
\end{equation}

and
\begin{equation}
    N_{\nu,k}=\frac{\sqrt{k!\,\Gamma(2\nu+2+k)}}{2^{\nu+1}\Gamma(\nu+k+\tfrac{3}{2})}
\end{equation}

\subsection{Quasinormal modes in Schwarzschild BTZ black hole}
\label{sec:btz}
The thermal bath that we are coupling our BPS-Janus system to is the BTZ black hole on the same geometry, on both sides of the interface. In this section we review the quasinormal modes of the BTZ black hole \cite{Birmingham_2001,Cardoso_2001}.

If we use the conventional cylinder coordinates for BTZ with $\ell=1$, $J=0$,
\begin{equation}
  ds^2
=
-(r^2-r_+^2)\,dt^2
+\frac{1}{r^2-r_+^2}\,dr^2
+r^2\,d\phi^2 ,
\qquad \phi\sim\phi+2\pi .
\end{equation}

The KG equation in BTZ for a scalar with dimension $\Delta_{BTZ}$ has solution $\Psi^{BTZ} = R(r)e^{-i\omega t}e^{i m\phi}$. Solving the radial equation gives
\begin{equation}
    R(r)=z^{-i\tfrac{\omega}{2r_+}}(1-z)^{\tfrac{2-\Delta_{BTZ}}{2}}\,_2F_1(a,b;c;z)
    \label{eq:radialbtz}
\end{equation}

with $z=1-\tfrac{r_+^2}{r^2}$, and $a,b=\tfrac{2-\Delta_{BTZ}}{2}+ i\tfrac{1}{2r_+}(-\omega\pm m)$, $c=1-i\tfrac{\omega}{r_+}$. Near the boundary $z\to 1$ or $r\to\infty$, we have expansion 
\begin{equation}
    \Psi_\omega^{BTZ}\sim \sum_m(\alpha_m^{BTZ}\, (r/r_+)^{-(2-\Delta_{BTZ})}+\beta_m^{BTZ}\, (r/r_+)^{-\Delta_{BTZ}})e^{im\phi}
    \label{eq:btzexpand}
\end{equation}

The normalizable and non-normalizable modes are

\begin{equation}
\begin{split}
    \alpha_m^{BTZ}(\omega,\Delta_{BTZ}) \;=\;\frac{\Gamma(\Delta_{BTZ})\,
\Gamma\!\left(1 - \tfrac{i \omega}{r_+}\right)}{(\Delta_{BTZ}-1)\,
\Gamma\!\left(\tfrac{\Delta_{BTZ} - i(\omega - m)/r_+}{2}\right)\,
\Gamma\!\left(\tfrac{\Delta_{BTZ} - i(m + \omega)/r_+}{2}\right)}
\\
\beta_m^{BTZ}(\omega,\Delta_{BTZ}) \;=\;\frac{\Gamma(1-\Delta_{BTZ})
\Gamma\!\left(1 - \tfrac{i \omega}{r_+}\right)}{
\Gamma\!\left(\tfrac{-\Delta_{BTZ} - i(\omega - m)/r_+ + 2}{2}\right)\,
\Gamma\!\left(\tfrac{-\Delta_{BTZ} - i(m + \omega)/r_+ + 2}{2}\right)}
\end{split}
\label{eq:bhab}
\end{equation}

The quasinormal modes of the BTZ black hole without rotation is then (after demanding that $\alpha$ vanishes) \cite{Cardoso_2001}
\begin{equation}
    \omega = \pm m-ir_+(\Delta_{BTZ}+2n)
\end{equation}

\subsection{Normal modes of BPS-Janus}
\label{sec:bps}
A canonical class of bulk geometries dual to interface CFTs with a coupling
jump is provided by Janus models \cite{Bak_2007,Bak_2011}.  Among them, a
two-parameter family of top-down supergravity solutions on
AdS$_2\times S^2\times T^4\rtimes\Sigma$ describes two-dimensional
$\mathcal{N}=(4,4)$ ICFTs with BPS interfaces \cite{Chiodaroli_2010}.  The
scalar Klein--Gordon equation and BPS-Janus normal modes were obtained in
\cite{Chiodaroli:2016jod}; we review the relevant results below.

The BPS-Janus geometry we consider here is a deformation of the type IIB supergravity on AdS$_3\times S^3\times T^4$, preserving
$SO(2,1)\times SO(3)$ isometries and eight supercharges \cite{Chiodaroli_2010}.
The boundary theory is a $2$d $\mathcal{N}=(4,4)$ D1/D5 CFT deformed by marginal couplings with a step-function
profile across the half-BPS interface. In particular, the six-dimensional dilaton and
a linear combination of the axion and the RR four-form interpolate between two asymptotic values on
opposite sides of the interface.

The ten-dimensional metric can be written as a fibration over a Riemann surface $\Sigma$ with boundary,
\begin{equation}
ds^2_{10}
= f_{1,10}^2\, ds^2_{AdS_2} + f_{2,10}^2\, ds^2_{S^2} + \rho_{10}^2\, dw\, d\bar w + f_{3,10}^2\, ds^2_{T^4},
\qquad w\in\Sigma,
\end{equation}
with all warp factors depending only on $w,\bar w$.
It is convenient to reduce to six dimensions (Einstein frame) and write
\begin{equation}
ds_6^2 = f_1^2\, ds^2_{AdS_2} + f_2^2\, ds^2_{S^2} + \rho^2\, dw\, d\bar w,
\label{eq:6dmetric}
\end{equation}
together with identities
\begin{equation}
f_1^2 f_2^2 = H^2,
\qquad
\sqrt{-g}=\rho^2 H^2,
\label{eq:Hidentity}
\end{equation}
where $H$ is one of the harmonic functions specifying the solution.

We focus on the simplest Janus solutions with two asymptotic AdS$_3$ regions and $\Sigma$ an infinite
strip. Writing
\begin{equation}
w = x + i y,
\qquad x\in\mathbb{R},\quad y\in(0,\pi),
\end{equation}
the two asymptotic AdS$_3$ regions are approached as $x\to\pm\infty$.
Nontrivial Janus deformations are parameterized by two real parameters $(\psi,\theta)$ controlling,
respectively, the dilaton and axion jumps; the undeformed limit is $\psi=\theta=0$.
A convenient combination is the interface parameter
\begin{equation}
\kappa \equiv \cosh\psi\,\cosh\theta.
\label{eq:kappa_def}
\end{equation}
For the RR-supported Janus family, the harmonic function and the ratios of metric factors take the
simple form
\begin{equation}
H = 2\, \cosh x\, \sin y,
\qquad
\frac{\rho^2}{f_1^2} = \frac{\kappa^2}{\cosh^2 x},
\qquad
\frac{\rho^2}{f_2^2}
= \frac{1}{\sin^2 y} + \frac{\kappa^2-1}{\cosh^2 x}.
\label{eq:ratios}
\end{equation}

Following \cite{Chiodaroli:2016jod} we now consider a massless scalar
propagating in the six-dimensional BPS Janus background. The Klein--Gordon (KG) equation may be written as
\begin{equation}
\Box_6\Phi = 0,
\end{equation}
or, using the metric \eqref{eq:6dmetric} and the identity \eqref{eq:Hidentity},
\begin{equation}
\frac{1}{\rho^2}\left[
\frac{\partial_a\!\left(H^2\partial_a\right)}{H^2}
+ \frac{\rho^2}{f_1^2}\nabla^2_{AdS_2}
+ \frac{\rho^2}{f_2^2}\nabla^2_{S^2}
\right]\Phi
 = 0,
\qquad a=x,y .
\label{eq:KGmaster_Delta}
\end{equation}

The solution is fully separable
\begin{equation}
    \Psi_\omega=f_{Ln}(x)g_{knLl}(\sigma)e^{-i\omega t}Y_{Ll}(y,\theta,\phi)
    \label{eq:solution}
\end{equation}

with
\begin{equation}
\nabla^2_{AdS_2} g = \hat\nu(\hat\nu+1) g,
\qquad
\nabla^2_{S^2} Y = -l(l+1) Y.
\end{equation}

Using the explicit Janus metric ratios \eqref{eq:ratios}, the remaining $(x,y)$ equation becomes
\begin{equation}
\left[
\partial_x\!\left(\cosh^2 x\,\partial_x\right)
+\partial_y\!\left(\sin^2 y\,\partial_y\right)
-\frac{l(l+1)}{\sin^2 y}
+\frac{\nu(\nu+1)}{\cosh^2 x}
\right]\Psi_{\hat\nu l}(x,y)=0,
\label{eq:xy_Delta}
\end{equation}
and regularity demands the quantization of parameters
\begin{equation}
\nu_n(\nu_n+1)\equiv (\Delta+n-1)(\Delta+n)
=
l(l+1)
+\kappa^2\!\left[\hat\nu_n(\hat\nu_n+1)-l(l+1)\right].
\end{equation}

The $L$ quantum number is from the Laplacian of the three-sphere of the compact $y$ direction and $S^2$. Reducing to AdS$_3$/CFT$_2$, the dimension of this operator in the dual 2D CFT is then $\Delta=2+L$. Since we will add the double-trace deformation using this operator and another CFT operator, we fix $L=l=0$ and keep the dimension of the scalar to be $\Delta=2$ onward. In presence of the SUSY Janus defect, imposing the absence of sources at the asymptotic AdS$_3$ boundaries quantizes the allowed values
of $\nu$, producing a discrete BOPE tower $\{\hat\Delta_n=\hat\nu_n+1\}$ (and corresponding normal modes) in the presence of the interface. In the AdS$_2$ part we simply do the substitutions $\Delta_n\to\hat{\Delta}_n,\nu_n\to\hat{\nu}_n$ in \eqref{eq:ads2} where $\nu(\nu+1)=\kappa^2\hat{\nu}_n(\hat{\nu}_n+1)$. Hence, the massless BOPE primary dimensions for the BPS-Janus interface are
\begin{equation}
\hat{\Delta}_n^{(\kappa)}
  =\frac12+
    \sqrt{\frac14+\frac{(n+1)(n+2)}{\kappa^2}}.
  \label{eq:bope-tower}
\end{equation}

In general, normal modes of the scalar operator can be seen from its BOPE tower on a (topological or not) defect, after  Fourier transforming the position space Green's function \cite{Aharony_2003}. Let us first do Euclidean and in the end analytic continue.
Transform from Poinc\'are $(x_\perp,\tau_P)$ to cylinder $(\phi,\tau)$:
\begin{equation}
    x_\perp=\frac{\sin\phi}{\cosh \tau+\cos\phi},\qquad 
\tau_{\rm P}=\frac{\sinh \tau}{\cosh \tau+\cos\phi}.
\end{equation}

The cross ratio is then
\begin{equation}
    \xi=\frac{\cosh(\tau-\tau')-\cos(\phi-\phi')}{2\,\sin\phi\,\sin\phi'},
\qquad
2\xi+1=\frac{\cosh(\tau-\tau')-\cos\phi\cos\phi'}{\sin\phi\,\sin\phi'}.
\end{equation}

For a scalar of dimension $\Delta$ with defect BOPE tower
$\hat{\Delta}_n$, the position-space Green function is
\begin{equation}
\begin{aligned}
  G_E(\phi,\tau;\phi',\tau')
  ={}&\frac{2^{\,3-2\Delta}}
          {\pi\,\kappa^{2\Delta}[\Gamma(\Delta-1)]^{2}}
      \frac{1}{|\sin\phi\,\sin\phi'|^{\Delta}}
      \sum_{n=0}^{\infty}\operatorname{sign}(\xi)^{\,n}
      \Big(n+\Delta-\tfrac12\Big)
  \\
  &\times\frac{\Gamma(n+2\Delta-1)}{\Gamma(n+1)}
      Q_{\hat\Delta_n-1}\!\big(|2\xi+1|\big).
\end{aligned}
\end{equation}

The Legendre function can be expanded as
\begin{equation}
    Q_{\hat\Delta_n-1}(\chi)=\sum_{s=0}^{\infty}a_s\,\chi^{-(\hat\Delta_n+2s)},\qquad
a_s=\frac{\sqrt\pi\,\Gamma(\hat\Delta_n+2s)}{2^{\hat\Delta_n+2s}\,s!\,\Gamma(\hat\Delta_n+\tfrac12+s)} .
\label{eq:Qexp}
\end{equation}

We now Fourier transform this into momentum space. Note that since defect breaks the translational invariance on the space $\phi$, the Green's function is not diagonal in the plane wave basis. 

We use the orthogonal basis $\Psi=e^{-i\omega t}f_k(\phi)$ from the AdS$_2$ Gegenbauer eigenfunctions in order to diagonalize the Green's function 
\begin{equation}
    \psi^{(n)}_k(\phi)=(\sin\phi)^{\hat\Delta_n}\,C^{(\hat\Delta_n)}_k(\cos\phi),\qquad k=0,1,2,\dots
\end{equation}

with inner products
\begin{equation}
    \|\psi^{(n)}_k\|^{2}=\int_0^\pi \big(\psi^{(n)}_k(\phi)\big)^2 d\phi
=\frac{\pi\,2^{\,1-2\hat\Delta_n}\,\Gamma(k+2\hat\Delta_n)}{k!\,(k+\hat\Delta_n)\,[\Gamma(\hat\Delta_n)]^{2}} .
\end{equation}

Notice that from now on $\phi$ ranges in $[0,\pi]$. It is half of the cylinder. It is no coincidence that the Fourier transform of $Q$ is exactly the momentum G function in this basis. 

from the expansion \eqref{eq:Qexp} and the Gegenbauer addition theorem
we get the components of the Euclidean Green's function:
\begin{equation}
    \int d\tau\,e^{-i\omega_E\tau}\,Q_{\hat\Delta_n-1}(\chi)
=2\pi\sum_{k=0}^{\infty}\frac{1}{\|\psi^{(n)}_k\|^{2}}\,\frac{\psi^{(n)}_k(\phi)\,\psi^{(n)}_k(\phi')}
{\omega_E^{2}+(\hat\Delta_n+k)^{2}}.
\end{equation}

In the end in Lorentzian we have
\begin{equation}
G_R(\omega;\phi,\phi')=\sum_{n=0}^{\infty}\sum_{k=0}^{\infty}
\mathcal R_{n,k}\;
\frac{\Phi_{n,k}(\phi)\Phi_{n,k}(\phi')}
{(\omega+i\epsilon)^{2}-(\hat\Delta_n+k)^{2}}
\label{eq:greenjanus}
\end{equation}
 where the modified basis are
\begin{equation}
    \Phi_{n,k}(\phi)=(\sin\phi)^{\hat{\Delta}_{n}-\Delta}\,C^{(\hat{\Delta}_{n})}_{k}(\cos\phi)
\end{equation}
and constants
\begin{equation}
    \mathcal R_{n,k}=\frac{2^{\,3+2(\hat{\Delta}_n-\Delta)}}{\pi\,\kappa^{2\Delta}}\;
\frac{\big(n+\Delta-\tfrac12\big)\,\Gamma(n+2\Delta-1)}{n!\,[\Gamma(\Delta-1)]^{2}}\;
\frac{k!\,(k+\hat\Delta_n)\,[\Gamma(\hat\Delta_n)]^{2}}{\Gamma(k+2\hat\Delta_n)}.
\label{eq:rnkj}
\end{equation}

In terms of a fixed defect channel mode $k$ we have
\begin{equation}
    G_R(\omega,k)=\sum_{n=0}^{\infty}\frac{\mathcal R_{n,k}}{(\omega+i\epsilon)^{2}-(\hat\Delta_n+k)^{2}}.
    \label{eq:greenfunctionwm}
\end{equation}

The normal modes are then $\omega_{n,k}=\hat{\Delta}_n+k$. Plugging in the massless scalar $\Delta=2$ and its BOPE tower \eqref{eq:bope-tower} in presence of the BPS-Janus interface, its normal modes are then
\begin{equation}
\Omega_{n,k}^{(\kappa)}=k+\hat{\Delta}_n^{(\kappa)}
  =k+\frac12+
    \sqrt{\frac14+\frac{(n+1)(n+2)}{\kappa^2}}.
  \label{eq:massless-Janus-normal-modes}
\end{equation}

\section{Coupling BPS-Janus to BTZ bath}
\label{sec:couple}

\subsection{Double trace coupling}

In this section we introduce dissipation by coupling the interface CFT to an
external thermal bath CFT through a double-trace deformation.  In the bulk we
consider two separate spacetimes, a Janus geometry and a BTZ black hole, whose
fields communicate through boundary conditions on the coupled half-space.
From the CFT perspective, the zero-temperature ICFT and the thermal bath
exchange energy on only one side of the interface.

The interaction begins with two decoupled CFTs, a ``system'' and a ``bath'',
and adds the marginal deformation
\cite{Aharony:2001pa,Witten:2001ua,Aharony:2006hz,Kiritsis:2006hy,giombi2025rginterfacesdoubletracedeformations,Karch:2023wui,Hartman:2006dy}
\begin{equation}
S_{\text{int}} =  \int d^d x \, h(x) \mathcal{O}_{\text{sys}}(x)\,
\mathcal{O}_{\text{bath}}(x),
\end{equation}
with $\Delta_{\text{sys}}+\Delta_{\text{bath}}=d$.  Here the system is the
Janus ICFT, so $\Delta_{\text{sys}}=\Delta_J$, and the bath is the BTZ theory,
so $\Delta_{\text{bath}}=\Delta_{B}$.
In the bulk, this interaction is implemented by transparent boundary conditions
relating the normalizable and non-normalizable modes of dual bulk fields across
two asymptotically AdS spacetimes.
Although the combined system evolves unitarily, the subsystem corresponding to
the original CFT exhibits dissipative behavior, manifested through
energy fluxes and the appearance of quasinormal modes with nonzero imaginary
frequencies.

Two natural coupling profiles $h(x)$ are a uniform boundary half-space and a point
contact.  The latter most directly parallels the free-fermion construction of
\cite{barad2025dissipationmeetsconformalinterface}.  A previous version of this
paper used a point source and followed a high-frequency branch as a candidate
minimum relaxation gap.  Although the source Fourier profile was fixed
correctly, the reciprocal condition was closed using only the zero Fourier
component of the response rather than its full value at the contact.  The
corrected point interaction also produces a rank-one self-energy with dark and
subradiant sectors.  Those earlier numerical results therefore do not 
establish a nonzero Liouvillian gap and are not used here.  We focus
exclusively on the uniform half-space coupling.

One additional subtlety arises from the top-down nature of the model.  The
separable supergravity fluctuation analyzed in the BPS Janus background is a
massless scalar, dual in standard quantization to a dimension-two operator
$\mathcal O_{\text{sys}}$ in the D1/D5 ICFT \cite{Chiodaroli:2016jod}.  A classically
marginal half-space interaction would therefore require a field of dimension
 zero on the bath side.  Furthermore, the near-boundary expansion of a
 massless scalar includes logarithmic terms signaling running local kernels in
 the corresponding double-trace interaction.  Rather than interpreting the
 dimension-zero field as an ordinary conformal primary, which would leave only
 the identity operator, we use the dynamical-source interpretation
of the endpoint alternate quantization.

More precisely, consider a second D1/D5 theory in a thermal state with a
dimension-two scalar operator $\mathcal O_{\text{bath}}$, again dual to a
massless bulk field.  Let $X_{\text{bath}}$ be its source.  We promote this
source to a dynamical boundary field by performing the endpoint Legendre transform, or
equivalently by imposing the alternate endpoint boundary condition on the
massless BTZ scalar \cite{Klebanov:1999tb,Witten:2001ua,Hartman:2006dy}.  The
two theories are then coupled over the boundary half-space $\mathcal R$ through
\begin{equation}
    \delta S_{\rm int}
    =
    -h_R\int_{\mathcal R} d^2x\,
    X_{\text{bath}}(x)\,\mathcal O_{\text{sys}}(x).
    \label{eq:dynamical-source-coupling}
\end{equation}
The field $X_{\text{bath}}$ has engineering dimension zero because it is the source for a
dimension-two operator in two dimensions.  It need not itself be an ordinary
local conformal primary.

The logarithmic behavior of the massless endpoint has a direct interpretation
in this language.  Integrating out the thermal bath D1/D5 theory at fixed
$X_{\text{bath}}$ generates a quadratic effective action for the dynamical
source,
\begin{equation}
    W_B[X_{\text{bath}}]
    =
    \frac12
    \int\frac{d^2p}{(2\pi)^2}\,
    X_\text{bath}(-p)\,\Pi_B(p;\mu)\,X_{\text{bath}}(p)+\cdots ,
\end{equation}
\begin{equation}
    \Pi_B(p;\mu)
    =
    C_B p^2\log\!\left(\frac{p^2}{\mu^2}\right)
    +Z_B(\mu)p^2+\cdots .
    \label{eq:induced-source-kernel}
\end{equation}
Here \(C_B\) is fixed by the bath two-point-function normalization, while
\(Z_B(\mu)\) is the renormalized coefficient of the local kinetic term.
Thus the dynamical-source propagator behaves schematically as
\begin{equation}
    G_X(p)
    \sim
    \frac{1}{
    p^2\left[
    Z_B(\mu)+C_B\log(p^2/\mu^2)
    \right]} .
\end{equation}
At finite temperature, the logarithm is replaced by the corresponding thermal
digamma functions appearing in the massless BTZ response.  The logarithmic
Fefferman--Graham terms therefore encode the induced kinetic kernel and its
renormalization-scale dependence \cite{Bianchi:2001kw,Skenderis:2002wp}.
The interaction in \eqref{eq:dynamical-source-coupling} is classically
marginal, but the endpoint logarithms generate running local boundary kernels;
in particular, a purely off-diagonal constant coupling is not by itself closed
under changes of subtraction scale.

The reciprocal mixed boundary conditions used below are the holographic
implementation of this dynamical-source coupling.  A literal identification of
$X$ with a decoupled free scalar or singleton requires a more specific endpoint
limiting prescription \cite{Bashmakov:2016wxk,Ohl:2012bk}; this additional
interpretation will not be needed for the spectral analysis carried out here.
The logarithms make finite projected damping parameters depend on the chosen
renormalized boundary kernel, which must be fixed before any UV comparison is
made.

Below we relax normalizability at the coupled Janus face and determine the
source and renormalized response profiles.  We restrict to the internal
singlet $L=l=0$, for which the separated Janus equation is known analytically.
The Janus solution is naturally expanded in AdS$_2$ descendant channels,
whereas the BTZ response is diagonal in full-cylinder Fourier momentum.  After
transforming the Janus asymptotics to the common cylinder Fefferman--Graham
frame, the reciprocal boundary conditions therefore give an infinite operator
problem.  We do not solve its complete quasinormal-mode spectrum. Instead, we evaluate
only the fixed-$k$, large-radial-level projected scalar channel of Section
\ref{sec:cuv}.

\subsection{Massless Fefferman-Graham}
\label{sec:massless-janus-data}

We now specialize to the massless six-dimensional scalar in the internal
singlet sector $L=l=0$.  This is the top-down fluctuation for which the
separated BPS-Janus equation is known.  Section \ref{sec:normal-modes} derived
its normal modes; here we need the corresponding wavefunctions and endpoint
data.  At fixed frequency
$\omega$ the scalar eigenfunction reads
\begin{equation}
  \Psi^{J}_{\omega}(x,\sigma,t)
  = e^{-i\omega t}
    \sum_{k\geq 0}J_k\,f_{\nu_k}(x)\,g_{\omega,k}(\sigma),
  \label{eq:massless-Janus-decomposition}
\end{equation}
where
\begin{equation}
  \nu_k
  =-\frac{1}{2}+\sqrt{\kappa^2(\omega-k-1)(\omega-k)+\frac{1}{4}}
  \label{eq:massless-separation-relation}
\end{equation}
and
\begin{equation}
  g_{\omega,k}(\sigma)
  =
  \frac{\sqrt{k!\,\Gamma(2\omega-k)}}
       {2^{\omega-k}\Gamma(\omega+\tfrac12)}
  (\cos\sigma)^{\omega-k}
  P_k^{(\omega-k-\frac12,\,\omega-k-\frac12)}(\sin\sigma).
  \label{eq:massless-ads2-wavefunction}
\end{equation}
For complex frequency, the square-root branch is chosen by analytic
continuation from the closed pole \(\nu_k=n+1\). 
The radial solution regular in the interior and normalizable at the uncoupled boundary $x\to\infty$ is
\begin{equation}
  f_\nu(x)
  =\frac{C_\nu}{2\cosh x}
    P^{-1}_{-\nu-1}(\tanh x),
  \qquad
  C_\nu^2=2\nu(\nu+1)(2\nu+1).
  \label{eq:massless-Janus-radial-solution}
\end{equation}

At the coupled face $x\to-\infty$, the two massless falloffs differ by an
integer and the Fefferman--Graham expansion contains a logarithm.  In the
normalization of \eqref{eq:massless-Janus-radial-solution},
\begin{equation}
  f_\nu(x)
  =\alpha_\nu
   +e^{2x}\left[
      \beta_{\nu,R}
      -2\nu(\nu+1)\alpha_\nu x
    \right]
   +\mathcal O(xe^{4x}),
  \label{eq:massless-Janus-x-expansion}
\end{equation}
where
\begin{equation}
  \alpha_\nu
  =\frac{C_\nu}{\Gamma(1-\nu)\Gamma(\nu+2)}
  =\frac{C_\nu\sin(\pi\nu)}{\pi\nu(\nu+1)}
  \label{eq:massless-Janus-alpha}
\end{equation}
and
\begin{equation}
  \beta_{\nu,R}
  =-\alpha_\nu\left\{
     1+\nu(\nu+1)
     \left[
       \psi(\nu+2)+\psi(1-\nu)-\psi(1)-\psi(2)
     \right]
   \right\}.
  \label{eq:massless-Janus-beta}
\end{equation}
Here $\psi(z)=\Gamma'(z)/\Gamma(z)$ is the digamma function. It is useful to use the equivalent expression
\begin{align}
  \beta_{\nu,R}
  ={}&-\alpha_\nu\left\{
     1+\nu(\nu+1)
     \left[
       \psi(\nu+2)+\psi(\nu)-\psi(1)-\psi(2)
     \right]
   \right\}
   -C_\nu\cos(\pi\nu).
  \label{eq:massless-Janus-beta-stable}
\end{align}
In particular,
\begin{equation}
  \alpha_N=0,
  \qquad
  \lim_{\nu\to N}\beta_{\nu,R}=(-1)^{N+1}C_N,
  \qquad N\in\mathbb Z_{>0}.
  \label{eq:massless-Janus-integer-limit}
\end{equation}
The source-free normal modes therefore obey \(\nu=n+1\), recovering the BOPE
tower in \eqref{eq:bope-tower}.

To impose the mixed boundary conditions, the Janus and BTZ fields must be
written in the same cylinder Fefferman--Graham frame.  Near the coupled face,
\begin{equation}
  e^x=\frac{1}{2\kappa r\cos\varphi}+\mathcal O(r^{-3}),
  \qquad
  \sigma=-\varphi+\mathcal O(r^{-2}).
  \label{eq:Janus-cylinder-map-massless}
\end{equation}
We parameterize the resulting Janus expansion as
\begin{equation}
  \Phi^J
  =a^J(\varphi)
   +\frac1{r^2}\left[
       b_{R}^J(\varphi;\mu_J)
       +\Gamma^J[a^J](\varphi)
        \log\!\left(\frac r{\mu_J}\right)
     \right]
   +\cdots .
  \label{eq:Janus-cylinder-FG}
\end{equation}
Here and below the subscript \(R\) on an asymptotic coefficient means
``renormalized'', and the coupled region is denoted by
\(\mathcal R\).
For one separated channel, with abbreviation
$g_k(\varphi)\equiv g_{\omega,k}(-\varphi)$, from \eqref{eq:massless-Janus-x-expansion} we have
\begin{equation}
  a_{k}^J(\varphi)
  =\alpha_{\nu_k}\,g_k(\varphi),
  \label{eq:Janus-cylinder-source}
\end{equation}
\begin{equation}
  \Gamma_{k}^J(\varphi)
  =\frac{2\nu_k(\nu_k+1)\alpha_{\nu_k}}
         {(2\kappa\cos\varphi)^2}
    g_k(\varphi),
  \label{eq:Janus-cylinder-log}
\end{equation}
and
\begin{align}
  b_{k,R}^J(\varphi;\mu_J)
  ={}&\frac{g_k(\varphi)}{(2\kappa\cos\varphi)^2}
     \left[
       \beta_{\nu_k,R}
       +2\nu_k(\nu_k+1)\alpha_{\nu_k}
        \log\!\bigl(2\kappa\mu_J\cos\varphi\bigr)
      \right]
     +\mathcal L_{\rm loc}[a_{k}^J](\varphi).
  \label{eq:Janus-cylinder-response}
\end{align}
Here \(\mathcal L_{\rm loc}\) is a scheme-dependent local linear operator
acting on \(a_k^J\).  It collects the \(O(r^{-2})\) terms generated by the
subleading Fefferman--Graham coordinate transformation together with finite
local counterterms.  Because it is linear in the source,
\(\mathcal L_{\rm loc}[a_k^J]\) vanishes at each normalmode pole of the system where
\(a_k^J=0\).  Below when we compare the boundary condition of BPS-Janus and the reference $\kappa=1$ case, we fix coupling \(h_R\), the subtraction scales \((\mu_J,\mu_B)\), and the coefficients
of all finite local terms. The scale dependence of the full kernel will be made explicit in
Section~\ref{sec:massless-running-kernel}.

We next do the Fefferman-Graham expansion for the ingoing massless scalar in BTZ.

For frequency $\omega$ and $U(1)$ momentum $m$ in the BTZ solution, let us define
\begin{equation}
  p_m=-\frac{i(\omega-m)}{2r_+},
  \qquad
  q_m=-\frac{i(\omega+m)}{2r_+},
  \label{eq:massless-BTZ-pq}
\end{equation}
and
\begin{equation}
  D_m(\omega;\mu_B)
  =\psi(1+p_m)+\psi(1+q_m)-\psi(1)-\psi(2)
   -2\log\!\left(\frac{\mu_B}{r_+}\right).
  \label{eq:massless-BTZ-D}
\end{equation}
The near-boundary expansion of the radial solution \eqref{eq:radialbtz} is
\begin{equation}
  R_m(r)
  =b_m^B
   +\frac1{r^2}\left[
      a_{m,R}^B(\mu_B)
      +\ell_m^B\log\!\left(\frac r{\mu_B}\right)
    \right]
   +\mathcal O(r^{-4}\log r),
  \label{eq:massless-BTZ-expansion}
\end{equation}
with
\begin{equation}
  b_m^B
  =\frac{\Gamma(1+p_m+q_m)}
         {\Gamma(1+p_m)\Gamma(1+q_m)},
  \label{eq:massless-BTZ-b}
\end{equation}
\begin{equation}
  \ell_m^B
  =-2r_+^2p_mq_m b_m^B
  =\frac{\omega^2-m^2}{2}\,b_m^B,
  \label{eq:massless-BTZ-log-coefficient}
\end{equation}
and
\begin{equation}
  a_{m,R}^B(\mu_B)
  =r_+^2b_m^B\left[
      p_mq_mD_m(\omega;\mu_B)
      -\frac{p_m+q_m}{2}
    \right].
  \label{eq:massless-BTZ-a}
\end{equation}
In the alternate quantization, $a_{m,R}^B$ is the source conjugate
to the dynamical variable $b_m^B$.  With our normalization, the retarded
response is
\begin{equation}
  G^B_{R,-}(\omega,m;\mu_B)
  =-2\frac{b_m^B}{a_{m,R}^B}
  =\frac{8}
  {(\omega^2-m^2)D_m(\omega;\mu_B)-2ir_+\omega}.
  \label{eq:massless-BTZ-Gminus}
\end{equation}
As emphasized before, the field with $\Delta_B=0$ should be understood as a dynamical source to the dimension-2 operator. 

\subsection{Uniform coupling over the boundary half-space}
\label{sec:massless-halfspace-matrix}

To carry out the uniform half-space coupling we still need to transform between the half- and full-cylinder eigenfunctions.  Let
\begin{equation}
  \mathcal S=S^1=(-\pi,\pi],
  \qquad
  \mathcal R=\left(-\frac\pi2,\frac\pi2\right)\subset\mathcal S
  \label{eq:halfspace-region}
\end{equation}
denote the full bath cylinder and the coupled boundary half-cylinder.  At
fixed frequency let
\begin{equation}
  P_{\mathcal R}:L^2(\mathcal S)\longrightarrow L^2(\mathcal R)
\end{equation}
be the restriction map to $\mathcal R$, and we can also pullback functions by
\begin{equation}
  (P_{\mathcal R}^{*}f)(\varphi)
  =
  \begin{cases}
    f(\varphi),&\varphi\in\mathcal R,\\
    0,&\varphi\in\mathcal S\setminus\mathcal R.
  \end{cases}
  \label{eq:halfspace-restriction-extension}
\end{equation}
Our normalization conventions for the source-response action are
\begin{equation}
  \delta S_J^{\rm ren}
  =\int_{\mathcal R}2b_R^J\,\delta a^J,
  \qquad
  \delta S_B^{\rm alt}
  =\int_{\mathcal S}(-2b^B)\,\delta a_R^B.
  \label{eq:massless-variation-normalization}
\end{equation}
Thus \(\mathcal O_{\rm sys}=2b_R^J\) and
\(X_{\rm bath}=-2b^B\).  Varying
\eqref{eq:dynamical-source-coupling} gives the reciprocal boundary conditions
\begin{equation}
  a^J=-2h_RP_{\mathcal R}b^B,
  \qquad
  a_{R}^B=2h_RP_{\mathcal R}^{*}b_{R}^J.
  \label{eq:massless-mixed-boundary-conditions}
\end{equation}

It is useful to collect \eqref{eq:massless-mixed-boundary-conditions} into a matrix boundary kernel form
\begin{equation}
  \begin{pmatrix}a^J\\ a_{R}^B\end{pmatrix}
  =\mathsf F_R
   \begin{pmatrix}b_{R}^J\\ b^B\end{pmatrix},
  \qquad
  \mathsf F_R=
  \begin{pmatrix}
    0&-2h_RP_{\mathcal R}\\
    2h_RP_{\mathcal R}^{*}&0
  \end{pmatrix}.
  \label{eq:massless-reference-boundary-kernel}
\end{equation}
The retarded responses are
\begin{equation}
  b_{R}^J=\frac12G^J_{R}\,a^J,
  \qquad
  b^B=-\frac12G^B_{R,-}\,a_{R}^B.
  \label{eq:massless-G-normalization}
\end{equation}
%Here the Green functions are response operators rather than pointwise quotients.

Eliminating the response data gives the homogeneous double-trace pole equation
\begin{equation}
 \left[
   \mathbf 1_{\mathcal R}
    -h_R^2\,G^B_{R,\mathcal R}\,G^J_{R}
 \right]a^J=0,
  \label{eq:massless-exact-operator-pole}
\end{equation}
where
\(G^B_{R,\mathcal R}
  =P_{\mathcal R}G^B_{R,-}P_{\mathcal R}^{*}\).
A quasinormal pole is a frequency for which the operator in brackets has a
nontrivial kernel. Introducing a finite regulator, this is equivalent to a vanishing determinant.  Although \(G^B_{R,-}\) is diagonal in
full-cylinder momentum, \(G^B_{R,\mathcal R}\) is generally not diagonal in a
half-interval basis.

Choose the orthonormal real basis on $\mathcal R$ as
\begin{equation}
  u_0(\varphi)=\frac1{\sqrt\pi},
  \qquad
  u_r(\varphi)=\sqrt{\frac2\pi}
  \begin{cases}
    \cos(r\varphi),&r>0\ \text{even},\\
    \sin(r\varphi),&r\ \text{odd},
  \end{cases}
  \qquad r=1,2,\ldots,
  \label{eq:halfspace-orthonormal-basis}
\end{equation}
and the normalized full-cylinder modes
\begin{equation}
  e_m(\varphi)=\frac{e^{im\varphi}}{\sqrt{2\pi}},
  \qquad m\in\mathbb Z.
  \label{eq:full-cylinder-Fourier-basis}
\end{equation}
For ingoing BTZ amplitudes $B_m$ the boundary data are
\begin{equation}
  b^B(\varphi)=\sum_{m\in\mathbb Z}B_m b_m^B e_m(\varphi),
  \qquad
  a_{R}^B(\varphi)
  =\sum_{m\in\mathbb Z}B_m a_{m,R}^B e_m(\varphi).
  \label{eq:full-cylinder-BTZ-boundary-data}
\end{equation}
We can now transform between half-space modes, full-\(S^1\) modes, and separated
Janus profiles, labelled by \(r,m,j\), respectively.  Here \(j\) is the
AdS$_2$ descendant index denoted by \(k\) in
\eqref{eq:massless-Janus-decomposition}; the new symbol avoids conflict with
the fixed target channel used below.  Define the Janus source and response
projections
\begin{equation}
  A_{rj}(\omega)
  =\int_{\mathcal R}d\varphi\,
    u_r(\varphi)^*a_{j}^J(\varphi),
  \qquad
  V_{rj}(\omega)
  =\int_{\mathcal R}d\varphi\,
    u_r(\varphi)^*b_{j,R}^J(\varphi),
  \label{eq:Janus-projection-matrices}
\end{equation}
and the overlap matrix
\begin{equation}
  K_{rm}
  =\left\langle u_r,P_{\mathcal R}e_m\right\rangle_{L^2(\mathcal R)}
  =\int_{\mathcal R}d\varphi\,
    u_r(\varphi)^*e_m(\varphi).
  \label{eq:restriction-overlap-matrix}
\end{equation}
It is important to note that the ordinary integral defining \(V_{rj}\) converges only when
\(\operatorname{Re}(\omega-j)>1\), and \(b^J_{j,R}\) belongs to
\(L^2(\mathcal R)\) only when
\(\operatorname{Re}(\omega-j)>3/2\).  At fixed \(\omega\), these conditions
fail for sufficiently large \(j\).  In addition, the orthogonal support projector on the full cylinder is $\Pi_{\mathcal R}=P_{\mathcal R}^{*}P_{\mathcal R}$ where
\begin{equation}
  (\Pi_{\mathcal R})_{mm'}
  =\int_{\mathcal R}d\varphi\,e_m(\varphi)^*e_{m'}(\varphi)
  =
  \begin{cases}
    \frac12,&m=m',\\[3pt]
    \displaystyle
    \frac{\sin[(m'-m)\pi/2]}{\pi(m'-m)},&m\ne m'.
  \end{cases}
  \label{eq:full-cylinder-support-projector}
\end{equation}
Let $K$ represents restriction to $\mathcal{R}$ and $K^\dagger$ represents the extension. In the complete bases, $\Pi_{\mathcal R}=K^\dagger K$ and
$KK^\dagger=\mathbf 1_{\mathcal R}$. 
The diagonal bath matrices
\begin{equation}
  \mathsf D_b(\omega)=\operatorname{diag}_{m\in\mathbb Z}(b_m^B),
  \qquad
  \mathsf D_a(\omega)=\operatorname{diag}_{m\in\mathbb Z}(a_{m,R}^B).
  \label{eq:BTZ-diagonal-matrices}
\end{equation}
act only on these full-cylinder Fourier amplitudes.

In terms of the matrices defined above, for Janus amplitudes \(J_j\) and BTZ amplitudes \(B_m\), the regulated
homogeneous double-trace system \eqref{eq:massless-exact-operator-pole} is
\begin{equation}
  \begin{pmatrix}
    A(\omega) & 2h_RK \mathsf D_b(\omega)\\
    -2h_RK^\dagger V(\omega) & \mathsf D_a(\omega)
  \end{pmatrix}
  \begin{pmatrix}J\\ B\end{pmatrix}=0.
  \label{eq:massless-full-block-matrix}
\end{equation}
Its nontrivial solutions determine the regulated quasinormal frequencies.
Whenever \(\mathsf D_a\) is invertible, eliminating \(B\) gives
\begin{equation}
 \left[
 A(\omega)
    +4h_R^2K \mathsf D_b(\omega)\mathsf D_a(\omega)^{-1}K^\dagger V(\omega)
 \right]J=0,
  \label{eq:massless-halfspace-kernel}
\end{equation}

The corresponding matrix response operators are
\begin{equation}
  G^J_{R}=2VA^{-1},
  \qquad
  G^B_{R,\mathcal R}=-2K\mathsf D_b\mathsf D_a^{-1}K^\dagger,
  \label{eq:responses}
\end{equation}
They reproduce \eqref{eq:massless-exact-operator-pole} on the regulated
domain.  To solve this system numerically, one must specify descendant, Fourier, and near-boundary
regulators.  In the next section we circumvent this and instead perform a fixed-descendant,
large-radial-level scalar projection. In that limit, we will prove that the above matrix essnentially factorizes, and we can analytically extract a UV quantity that measures how the dissipation rate is suppressed by the interface.

\subsection{Subtraction scales and the running boundary kernel}
\label{sec:massless-running-kernel}

Before going into deriving the quasinormal modes we need to resolve the subtlety involving the running scales. The massless interaction is classically marginal, but the logarithmic
Fefferman--Graham terms make the renormalized boundary conditions
scale-dependent.  From \eqref{eq:Janus-cylinder-FG} and
\eqref{eq:massless-BTZ-expansion}, the (non-)renormalized coefficients are
\begin{equation}
  b_{R}^J(e^{\ell_J}\mu_J)
  =b_{R}^J(\mu_J)+\ell_J\,\Gamma^J[a^J],
  \label{eq:Janus-scale-running}
\end{equation}
\begin{equation}
  a_{R}^B(e^{\ell_B}\mu_B)
  =a_{R}^B(\mu_B)+\ell_B\,\Gamma^B[b^B],
  \label{eq:BTZ-scale-running}
\end{equation}
where, for the separated modes above,
\begin{equation}
  \Gamma_{k}^J[a^J]
  =\frac{2\nu_k(\nu_k+1)\alpha_{\nu_k}}
         {(2\kappa\cos\varphi)^2}g_k(\varphi),
  \qquad
  \Gamma_{m}^B[b^B]
  =\frac{\omega^2-m^2}{2}b_m^B.
  \label{eq:massless-log-functionals}
\end{equation}
To see explicitly why the off-diagonal kernel is not closed under scale
changes, let primed coefficients denote the shifted scales in
\eqref{eq:Janus-scale-running} and \eqref{eq:BTZ-scale-running}.  Rewriting the
same physical boundary condition in terms of these coefficients gives
\begin{equation}
 \begin{pmatrix}a^J\\ a_R^{B\,\prime}\end{pmatrix}
 =
 \begin{pmatrix}
   0&-2h_RP_{\mathcal R}\\
   2h_RP_{\mathcal R}^{*}&
   \ell_B\Gamma^B
   +4h_R^2\ell_JP_{\mathcal R}^{*}\Gamma^JP_{\mathcal R}
 \end{pmatrix}
 \begin{pmatrix}b_R^{J\,\prime}\\ b^B\end{pmatrix}.
 \label{eq:massless-running-boundary-kernel}
\end{equation}
Thus a relation that is purely off-diagonal at the original scales acquires a
local bath-to-bath diagonal block when the scales are varied.  Setting this generated
term back to zero would define a different renormalized boundary condition.  Hence, a physical calculation must fix the full renormalized kernel at a reference scale.  Equivalently,
as shown below, one can hold \(\mu_J,\mu_B\), all finite local terms, and
\(h_R\) fixed while retaining the logarithms explicitly.  The identical
prescription is used for Janus and for the \(\kappa=1\) reference geometry.

The finite-frequency problem remains the full matrix
\eqref{eq:massless-full-block-matrix}. We will show that, for
a fixed AdS$_2$ descendant channel and large radial/BOPE level, the double-trace matrix is
asymptotically diagonal after suitable basis transformation.  We will use the resulting projected branch to define the
radial UV diagnostic $c_{\rm UV}$ without numerically solving the full quasinormal mode of the coupled system.

\section{The radial UV dissipation $c_{\text{UV}}$}
\label{sec:cuv}

The exact double-trace equation \eqref{eq:massless-full-block-matrix} mixes
many closed BPS-Janus modes with different \((n,k)\) but nearby frequencies
\(\Omega_{n,k}^{(\kappa)}\).  Solving the full regulated matrix is therefore a
degenerate spectral problem. It is especially very subtle when $\kappa$ is a general non-rational number even in the large-$\Re\,\omega$ limit. Already in the trivial case $\kappa=1$, we need to do a highly degenerate perturbation theory to obtain the quasinormal modes as there are $O(\omega)$ number of $(n,k)$ that give the same Janus normal mode.  

In this section, we fix the AdS$_2$ descendant level \(k\)
and take the large-\(n\), large-frequency limit. Starting from the full double-trace equation \eqref{eq:massless-halfspace-kernel} and \eqref{eq:responses}, we eliminate the bath amplitudes and define a projected matrix $\widehat{\mathcal{M}}$ in the system's AdS$_2$-descendant levels that incorporates the bath coupling in this sector. We prove that $\widehat{\mathcal{M}}$ is asymptotically diagonal, so its diagonal elements essentially give its eigenvalues. It is worth emphasizing that we do not claim that the corresponding
sector factorizes from the complete set of $(n,k)$ normal modes.

Solving the projected equation, we derive the leading imaginary frequency
shift and a relative radial-UV diagnostic \(c_{\text{UV}}\), comparing BPS
Janus with the \(\kappa=1\) reference theory at fixed scales. This quantity measures how the holographic interface suppresses the dissipation rate of the zero-T system. As we will see below the projected
damping parameter tends to zero for every fixed \(k\). It is thus very interesting to establish the relation between this quantity and the relaxation quantity of interfaces studied in \cite{barad2025dissipationmeetsconformalinterface}.

Throughout this section the subtraction scales and the finite local
counterterms are held fixed as mentioned in Section \ref{sec:massless-running-kernel}.
The same prescription is
used for BPS-Janus and for pure AdS$_3$.
Concretely, at the chosen reference scales \((\mu_J,\mu_B)\) we impose the
double-trace boundary condition
\eqref{eq:massless-reference-boundary-kernel}, with
its diagonal local entries set to zero as part of the renormalization
condition.
In the large-\(\zeta\) projected responses these terms shift the
nonlogarithmic constants \(c_J^{(\kappa)}\) and \(c_B\) introduced in
\eqref{eq:uv-janus-log} and \eqref{eq:uv-bath-log}, and hence
\(C_h^{(\kappa)}\) below.  These constants also contain finite pieces from the
digamma functions and coordinate matching, so they are not just counterterm
coefficients.  Such shifts do not alter the coefficient of \(\log\zeta\) and
therefore do not affect \(c_{\text{UV}}/c\).

\subsection{Fixed-channel projection at large $\omega$}

In this subsection we start from \eqref{eq:massless-halfspace-kernel} and project into the fixed-$k$ large $\omega$ channel by first using the
full-cylinder Fourier representation, then applying an analytic left projection in
channel space, and in the end recovering the diagonal piece at \(j=k\). 

At fixed frequency \(\omega\), let \(j\) denote the AdS$_2$ descendant index
and define
\begin{equation}
  \lambda_j=\omega-j,
  \qquad
  \Lambda_j=\lambda_j(\lambda_j-1),
  \label{eq:uv-channel-definitions}
\end{equation}
The separation relation \eqref{eq:massless-separation-relation} is then
\begin{equation}
  \nu_j(\nu_j+1)=\kappa^2\Lambda_j,
\end{equation}
while the boundary profile satisfies
\begin{equation}
  g_j''(\varphi)
  +\left[\omega^2-\Lambda_j\sec^2\varphi\right]
   g_j(\varphi)=0.
  \label{eq:uv-ads2-equation}
\end{equation}

Note that the BTZ response is diagonal in full-cylinder momentum but its
restriction to $\mathcal R$ is not diagonal.  We therefore use the normalized full-cylinder modes
\eqref{eq:full-cylinder-Fourier-basis} and extend the Janus wavefunctions by zero
outside $\mathcal R$.  Define the
flat and weighted projections
\begin{equation}
  \mathcal A_{mj}
  =\int_{\mathcal R}d\varphi\,
    e_m^*(\varphi)g_j(\varphi),
  \qquad
  \mathcal B_{mj}
  =\int_{\mathcal R}d\varphi\,
    e_m^*(\varphi)\sec^2\varphi\,g_j(\varphi).
  \label{eq:uv-fourier-projections}
\end{equation}
The change from roman to calligraphic letters records this change of
representation and the removal of radial coefficients.  In
Section~\ref{sec:massless-halfspace-matrix}, \(A_{rj}\) and \(V_{rj}\) are the
half-interval projections of the full Janus source and response, whereas
\(\mathcal A_{mj}\) and \(\mathcal B_{mj}\) are the flat and
\(\sec^2\varphi\)-weighted full-cylinder Fourier projections of the wavefunction \(g_j\).  Since
\(a_j^J=\alpha_{\nu_j}g_j\), completeness of the two bases gives the source and response matrices in full $S^1$ basis
\begin{align}
  A_{rj}
  &=\alpha_{\nu_j}\sum_{m\in\mathbb Z}K_{rm}\mathcal A_{mj},
  \nonumber\\
  (K^\dagger V)_{mj}
  &=\int_{\mathcal R}d\varphi\,
    e_m^*(\varphi)b^J_{j,R}(\varphi;\mu_J)
  \nonumber\\
  &=\frac{\beta_{\nu_j,R}
        +2\nu_j(\nu_j+1)\alpha_{\nu_j}\log(2\kappa\mu_J)}
       {4\kappa^2}\,\mathcal B_{mj}
  \nonumber\\
  &\quad
   +\frac{\alpha_{\nu_j}\Lambda_j}{2}
    \int_{\mathcal R}d\varphi\,
      e_m^*(\varphi)\sec^2\varphi\,g_j(\varphi)
      \log(\cos\varphi)
  \nonumber\\
  &\quad
   +\int_{\mathcal R}d\varphi\,
      e_m^*(\varphi)\mathcal L_{\rm loc}
      [\alpha_{\nu_j}g_j](\varphi).
  \label{eq:section3-section4-Janus-dictionary}
\end{align}

Recall that the last line depends on the finite local counterterms that will be fixed in our computation. Multiplying \eqref{eq:uv-ads2-equation} by \(e_m^*\) and integrating twice by
parts gives their relation
\begin{equation}
  \mathcal B_{mj}
  =\frac{\omega^2-m^2}{\Lambda_j}\,
   \mathcal A_{mj}.
  \label{eq:uv-projection-identity}
\end{equation}

The discarded boundary contribution is
\(
 [e_m^*g_j'-(e_m^*)'g_j]_{-\pi/2}^{\pi/2}/\Lambda_j
\).
Near either endpoint,
\(g_j\sim(\cos\varphi)^{\lambda_j}\), so both \(g_j\) and
\(g_j'\) vanish when \(\operatorname{Re}\lambda_j>1\).  This condition is true \(\omega\to\infty\) for the target channel \(j=k=O(1)\), but it need not hold when \(j\sim\omega\).  On the other hand, we have the weighted orthogonality condition
\begin{equation}
  \int_{\mathcal R}d\varphi\,
  \sec^2\varphi\,g_j(\varphi)g_{j'}(\varphi)
  =\frac{\delta_{jj'}}{2\lambda_j-1}.
  \label{eq:uv-weighted-orthogonality}
\end{equation}
This is the analytic bilinear orthogonality relation continued in $\omega$.
Equivalently, Parseval's identity implies
\begin{equation}
  \sum_{m\in\mathbb Z}
  \mathcal B_{-m,j}\mathcal A_{mj'}
  =\frac{\delta_{jj'}}{2\lambda_j-1}.
  \label{eq:uv-left-inverse}
\end{equation}
Thus $(2\lambda_j-1)\mathcal B_{-m,j}$ provides a left inverse of the
flat projection matrix.

This projection identity 
simplifies the bath-response factor inside the Schur-reduced equation
\eqref{eq:massless-halfspace-kernel}.  Combining it with the diagonal BTZ
response \eqref{eq:massless-BTZ-Gminus}, the bath response function is
\begin{equation}
  G^B_{R,-}(\omega,m;\mu_B)\mathcal B_{mj}
  =\frac{8}{\Lambda_j}\,
    \mathcal C_m(\omega;\mu_B)\mathcal A_{mj},
  \label{eq:uv-massless-cancellation}
\end{equation}
where
\begin{equation}
  \mathcal C_m(\omega;\mu_B)
  =
  \left[
    D_m(\omega;\mu_B)
    -\frac{2ir_+\omega}{\omega^2-m^2}
  \right]^{-1}.
  \label{eq:uv-bath-symbol}
\end{equation}
On the bath side, after projecting into the full space basis, \eqref{eq:responses} implies
\begin{equation}
  (\mathsf D_b\mathsf D_a^{-1})_{mm}
  =-\frac12G^B_{R,-}(\omega,m;\mu_B)
  =-\frac{4}{\omega^2-m^2}\,\mathcal C_m(\omega;\mu_B).
  \label{eq:section3-section4-bath-dictionary}
\end{equation}

The factor \(\omega^2-m^2\) generated by the AdS$_2$ equation
cancels the same factor in the massless BTZ endpoint response. This allows us to analytically solve the eigenvalues up to leading order in this branch.

Define the canonically normalized projections
\begin{equation}
  \widehat{\mathcal A}_{mj}
  =\sqrt{2\lambda_j-1}\,\mathcal A_{mj},
  \qquad
  \widehat{\mathcal B}_{mj}
  =\sqrt{2\lambda_j-1}\,\mathcal B_{mj}.
  \label{eq:uv-canonical-projections}
\end{equation}
To isolate the residual AdS$_2$ channel-mixing part of the Schur-reduced bath term in \eqref{eq:massless-halfspace-kernel},
we strip off the column factor \(8/\Lambda_{j'}\) from
\eqref{eq:uv-massless-cancellation} and the Janus radial source--response
coefficient.  The only channel map needed below is its canonically normalized bath response,
\begin{equation}
  \widehat{\mathcal M}_{jj'}(\omega)
  \equiv\sum_{m\in\mathbb Z}
    \widehat{\mathcal B}_{-m,j}\mathcal C_m(\omega;\mu_B)
    \widehat{\mathcal A}_{mj'}.
  \label{eq:uv-normalized-geometric-map}
\end{equation}
For an input Janus channel \(j'\), \(\mathcal A_{mj'}\) gives its flat
full-cylinder Fourier coefficients, \(\mathcal C_m\) multiplies them by the
residual bath symbol, and
\((2\lambda_j-1)\mathcal B_{-m,j}\) projects the result back onto channel
\(j\).  Thus \(\widehat{\mathcal M}\) conveniently represents the
bath part of the Schur-reduced matrix in
\eqref{eq:massless-halfspace-kernel}. Its diagonal element is the constant bath symbol seen by the normalized channel $j=k$. Notice that $\widehat{\mathcal M}$ does not see the cancelled factors of $8/\Lambda_{j'}$  and $\omega^2-m^2$ in
\eqref{eq:uv-massless-cancellation}
, which makes our calculation in this projected limit universal.

Let us now study the projected root connected to the \(n\)th radial mode in the descendant
channel \(j=k\). Define complex numbers \(s,\zeta\) by
\begin{equation}
  \nu\equiv\nu_k=n+1+\frac{s}{2},
  \qquad
  \zeta=\omega-k-\frac12.
  \label{eq:uv-radial-variables}
\end{equation}
Here \(s=0\) reproduces the closed mode labelled by \((n,k)\), while the bath coupling
shifts \(s\) and \(\omega\), into the complex plane.  The separation relation becomes
\begin{equation}
  \left(\nu+\frac12\right)^2
  =\kappa^2\zeta^2+\frac{1-\kappa^2}{4}.
  \label{eq:uv-eta-zeta}
\end{equation}
At $s=0$,
\begin{equation}
  \zeta_n^{(\kappa)}
  =
  \frac{1}{\kappa}
  \sqrt{\left(n+\frac32\right)^2
        +\frac{\kappa^2-1}{4}},
  \qquad
  \Omega_{n,k}^{(\kappa)}
  =k+\frac12+\zeta_n^{(\kappa)},
  \label{eq:uv-closed-frequency}
\end{equation}
and hence
\begin{equation}
  \omega(s)
  =k+\frac12+\frac{1}{\kappa}
   \sqrt{\left(n+\frac32+\frac{s}{2}\right)^2
          +\frac{\kappa^2-1}{4}}.
\end{equation}
For fixed $\kappa$, expansion about $s=0$ gives
\begin{equation}
  \omega(s)-\Omega_{n,k}^{(\kappa)}
  =\frac{s}{2\kappa}
   +O\!\left(\frac{s}{n^2}\right)
   +O\!\left(\frac{s^2}{n^3}\right).
  \label{eq:uv-frequency-shift}
\end{equation}

At fixed $k$, the canonically normalized AdS$_2$ wavefunction
\begin{equation}
  \widehat g_k
  =\sqrt{2\lambda_k-1}\,g_k
\end{equation}
approaches a fixed harmonic-oscillator wavepacket under the rescaling
$Y=\sqrt{\zeta}\,\varphi$.  In particular, it is concentrated around
$\Delta\varphi=O(\zeta^{-1/2})$. Equivalently, its Fourier modes is focused near $|m|=O(\zeta^{1/2})\ll\omega$. This will lead to the normalized bath map $\widehat{\mathcal{M}}$ being almost diagonal.

 More precisely, one can show that the fixed-degree Jacobi polynomials obeys the following bound: for every fixed integer $p\geq1$ and every fixed set of channels
$j=k+O(1)$,
\begin{equation}
  \sum_{m\in\mathbb Z}(1+m^2)^p
  \left(
    |\widehat{\mathcal A}_{mj}|^2
    +|\widehat{\mathcal B}_{mj}|^2
  \right)
  \leq C_{j,p}\zeta^p.
  \label{eq:uv-fourier-moments}
\end{equation}
Here \(C_{j,p}\) is independent of \(\zeta\).
Indeed, once $\operatorname{Re}\lambda_j>p+2$, the zero extensions of
$\widehat g_j$ and $\sec^2\varphi\,\widehat g_j$ have vanishing endpoint
traces through the required order.  Using Parseval's identity and differentiating the
fixed-degree Jacobi expression gives the above $\zeta^p$ scaling.

Fix $0<\epsilon<1/2$ and set
$M_\zeta=\zeta^{1/2+\epsilon}$.  It follows that, for every $L>0$,
\begin{equation}
  \sum_{|m|>M_\zeta}
  \left(
    |\widehat{\mathcal A}_{mj}|^2
    +|\widehat{\mathcal B}_{mj}|^2
  \right)
  =O_{j,L,\epsilon}(\zeta^{-L}).
  \label{eq:uv-fourier-tail}
\end{equation}
In particular, the contribution from the region $|m|\sim\omega$ is negligible and we can safely use small $|m|/\omega$ expansion below.

For $|m|\leq M_\zeta$ one has $m/\omega\to0$, so the large-argument expansion
of the digamma functions and gives
\begin{equation}
  D_m(\omega;\mu_B)-D_0(\omega;\mu_B)
  =
  \log\left(1-\frac{m^2}{\omega^2}\right)
  +O\left(\frac{m^2}{\omega^3}
          +\frac{m^4}{\omega^4}\right).
  \label{eq:uv-symbol-flatness}
\end{equation}
Let \(\mathcal C_0=\mathcal C_{m=0}\). For simplicity, we assume that there is no bath pole that happens to be very close to a BPS-Janus pole. Concretely, this means that we impose an upper bound on $\mathcal C_m$: there are constants $a,b,C>0$ such that
\begin{equation}
  \left|\frac{\mathcal C_m}{\mathcal C_0}\right|
  \leq C\zeta^a(1+|m|)^b,
  \qquad m\in\mathbb Z.
  \label{eq:uv-tail-symbol-bound}
\end{equation}
Then from \eqref{eq:uv-fourier-tail} we have that for every \(L>0\),
\begin{equation}
  \sum_{|m|>M_\zeta}(1+|m|)^{2b}
  \left(
    |\widehat{\mathcal A}_{mk}|^2
    +|\widehat{\mathcal B}_{mk}|^2
  \right)
  =O_{k,L,\epsilon}(\zeta^{-L}).
  \label{eq:uv-weighted-fourier-tail}
\end{equation}
Combining \eqref{eq:uv-symbol-flatness} with the bound
\eqref{eq:uv-fourier-moments}, and using
\eqref{eq:uv-tail-symbol-bound} and
\eqref{eq:uv-weighted-fourier-tail} on the complement range, we have
\begin{equation}
  \sum_{m\in\mathbb Z}
  \left|\frac{\mathcal C_m-\mathcal C_0}{\mathcal C_0}\right|^2
  \left(
    |\widehat{\mathcal A}_{mk}|^2
    +|\widehat{\mathcal B}_{mk}|^2
  \right)
  =O\left(\frac{1}{\zeta^2\log^2\zeta}\right).
  \label{eq:uv-symbol-norm-bound}
\end{equation}

From this we use the identity
\eqref{eq:uv-left-inverse} and Cauchy-Schwarz inequality, the diagonal elements of the normalized bath signal matrix $\widehat{\mathcal{M}}$ at large $n$ is
\begin{equation}
  \widehat{\mathcal M}_{kk}
  =\mathcal C_0
   \left[1+O\!\left(\frac{1}{n\log n}\right)\right].
  \label{eq:uv-geometric-diagonal}
\end{equation}
Let us study the off-diagonal elements. For every fixed $q\geq1$, expanding in even powers of $m/\omega$, the term
$m^{2q}$ becomes the $2q$-th power of oscillator momentum in the fixed-$k$
large-$n$ limit.  It carries an additional factor of $n^{-q}/\log n$.
Then for $|q|\le k/2$ we have
\begin{equation}
  \frac{\widehat{\mathcal M}_{k,k\pm 2q}}
       {\widehat{\mathcal M}_{kk}}
  =O\left(\frac{1}{n^q\log n}\right),
  \label{eq:uv-mixing-suppression}
\end{equation}
while opposite parities do not mix.

In order to show that the eigenvalue of the matrix is equal to the diagonal element at leading order, we need to prove the upper bound of the squared sum of all off-diagonal elements. Let \(\delta\mathcal C\) be the
full-cylinder Fourier multiplier defined by
\(\delta\mathcal C\,e_m=(\mathcal C_m-\mathcal C_0)e_m\), set
\(W=\sec^2\varphi\), and define
\begin{equation}
  \langle f,g\rangle_W
  \equiv
  \int_{\mathcal R}d\varphi\,W(\varphi)f(\varphi)^*g(\varphi).
  \label{eq:uv-weighted-inner-product}
\end{equation}
At the real closed frequency \(s=0\), the canonical profiles are orthonormal
in \(L^2(\mathcal R,W\,d\varphi)\), and the nonconstant part of the matrix can
be written as
\begin{equation}
  \delta\widehat{\mathcal M}^{\rm geom}_{kj'}
  =\left\langle
    W^{-1}\delta\mathcal C^\dagger W\widehat g_k,
    \widehat g_{j'}
   \right\rangle_W .
  \label{eq:uv-bessel-representation}
\end{equation}
Moreover, Parseval and $W^{-1}\leq1$ give the bound
\begin{equation}
  \left\|W^{-1}\delta\mathcal C^\dagger
          W\widehat g_k\right\|_W^2
  \leq
  \sum_{m\in\mathbb Z}
    |\mathcal C_m-\mathcal C_0|^2
    |\widehat{\mathcal B}_{mk}|^2.
  \label{eq:uv-bessel-bridge}
\end{equation}
At $s=0$, consider the regular set
$\mathcal I_{\rm reg}(\omega)=
\{j\in\mathbb Z_{\geq0}:\lambda_j>1\}$. Bessel's inequality together with
\eqref{eq:uv-symbol-norm-bound}, \eqref{eq:uv-geometric-diagonal}, and
\eqref{eq:uv-bessel-bridge} then gives the upper bound
\begin{equation}
  \frac{
    \left(\displaystyle\sum_{\substack{j\in\mathcal I_{\rm reg}(\omega)\\j\ne k}}
      |\widehat{\mathcal M}_{kj}|^2\right)^{1/2}}
       {|\widehat{\mathcal M}_{kk}|}
  =O\left(\frac{1}{n\log n}\right).
  \label{eq:uv-complete-row-decoupling}
\end{equation}

From this, we can then analytically derive the projected quasinormal pole at the large-$n$ limit for fixed descendant level $k$.

\subsection{Dissipation rate in the large-$n$ projected branch}

Now let us derive the analytical expression of the pole of \eqref{eq:massless-exact-operator-pole} in this fixed-$k$ large-$n$ branch. We calculate its imaginary part $-\mathrm{Im}\,\omega_{n,k,\mathrm{proj}}^{(\kappa)}$ and study its leading behavior in this limit.

Near the $n$-th Janus pole for large $n$, the reflection identity
\begin{equation}
  \psi(1-\nu)=\psi(\nu)+\pi\cot(\pi\nu)
\end{equation}
and the boundary coefficients
\eqref{eq:massless-Janus-alpha}--\eqref{eq:massless-Janus-beta} give
\begin{equation}
  \frac{\beta_{\nu,R}}{\alpha_\nu}
  =
  -\nu(\nu+1)
  \left[
    2\log\nu+2\gamma_{\!E}-1
    +\pi\cot\left(\frac{\pi s}{2}\right)
  \right]
  +O(\nu).
  \label{eq:uv-janus-ratio}
\end{equation}
We define the projected Janus response in the target channel by the analytic
bilinear projection
\begin{equation}
  G^J_{R,\mathrm{proj}}(\omega)
  \equiv
  \frac{2(2\lambda_k-1)}{\alpha_{\nu_k}}
  \int_{\mathcal R}d\varphi\,
    g_k(\varphi)b^J_{k,R}(\varphi;\mu_J).
  \label{eq:uv-projected-Janus-definition}
\end{equation}
 Using
\eqref{eq:Janus-cylinder-response}  and
\eqref{eq:uv-weighted-orthogonality} (with the convention in
\eqref{eq:massless-G-normalization}), its expression is
\begin{equation}
  G^J_{R,\mathrm{proj}}
  =
  \frac{1}{2\kappa^2}
  \left[
    \frac{\beta_{\nu,R}}{\alpha_\nu}
    +2\nu(\nu+1)\log(2\kappa\mu_J)
  \right]
  +\text{local terms}
  +O(\zeta).
  \label{eq:uv-janus-projection}
\end{equation}
Recall $\frac{\nu(\nu+1)}{\kappa^2}
  =\zeta^2-\frac14$ and $\log\nu=\log(\kappa\zeta)+O(\zeta^{-1})$.  Absorbing finite local terms into a $\zeta$-independent constant $c_J^{(\kappa)}$ gives
\begin{equation}
  G^J_{R,\mathrm{proj}}
  =
  -\frac{\Lambda_k}{2}
  \left[
    L_J^{(\kappa)}(\zeta)
    +\pi\cot\left(\frac{\pi s}{2}\right)
  \right]
  +O(\zeta),
  \label{eq:uv-janus-response}
\end{equation}
with
\begin{equation}
  L_J^{(\kappa)}(\zeta)
  =
  2\log\left(\frac{\zeta}{\mu_J}\right)
  +c_J^{(\kappa)}.
  \label{eq:uv-janus-log}
\end{equation}

On the other hand, the normalized bath map $\widehat{\mathcal{M}}$ from the previous subsection is effectively a diagonal bath response function in this branch
\begin{align}
  G^{B,\mathrm{eff}}_{R,k}(\omega;\mu_B)
  &\equiv
  \frac{8}{\Lambda_k}\,\mathcal C_0(\omega;\mu_B)
  \nonumber=
  \frac{8}
       {\zeta^2\left[L_B(\zeta)-i\pi\right]}
  \left[1+O(\zeta^{-1})\right],
  \label{eq:uv-btz-response}
\end{align}
with
\begin{equation}
  L_B(\zeta)
  =
  2\log\left(\frac{\zeta}{\mu_B}\right)+c_B.
  \label{eq:uv-bath-log}
\end{equation}
Note that it agrees with the \(m=0\) Fourier response
\(G^B_{R,-}(\omega,0)=8\mathcal C_0/\omega^2\) up to
relative \(O(\zeta^{-1})\) corrections.
The real constants$c_B$ and the above $c_J^{(\kappa)}$ are scheme-dependent.

Now we can finally turn the double-trace equations \eqref{eq:massless-mixed-boundary-conditions}, or equivalently \eqref{eq:massless-running-boundary-kernel} in this limit into the
leading order rank-1 equation
\begin{align}
  0
  &=
  1-h_R^2
  G^J_{R,\mathrm{proj}}G^{B,\mathrm{eff}}_{R,k}
  +o(1)
  \nonumber\\
  &=
  1+4h_R^2
  \frac{
    L_J^{(\kappa)}(\zeta)
    +\pi\cot(\pi s/2)}
  {L_B(\zeta)-i\pi}
  +o(1).
  \label{eq:uv-projected-secular}
\end{align}
Denote $ \xi_h=\frac{1}{4h_R^2}$ and $ \mathcal L_h^{(\kappa)}(\zeta)
  =
  L_J^{(\kappa)}(\zeta)+\xi_h L_B(\zeta)$. Equation \eqref{eq:uv-projected-secular} becomes
\begin{equation}
  \pi\cot\left(\frac{\pi s}{2}\right)
  =
  -\mathcal L_h^{(\kappa)}(\zeta)
  +i\pi\xi_h+o(1).
  \label{eq:uv-cot-equation}
\end{equation}
Using
$\pi\cot(\pi s/2)=2/s+O(s)$ the solution is
\begin{equation}
  s
  =
  -\frac{
    2\left[
      \mathcal L_h^{(\kappa)}(\zeta)+i\pi\xi_h
    \right]}
  {
    \left[\mathcal L_h^{(\kappa)}(\zeta)\right]^2
    +\pi^2\xi_h^2
  }
  +o(\log^{-2}\zeta).
  \label{eq:uv-s-shift}
\end{equation}
 We define the projected
frequency and dissipation rate by
\begin{equation}
  \gamma_{n,k,\mathrm{proj}}^{(\kappa)}
  \equiv
  -\operatorname{Im}\omega_{n,k,\mathrm{proj}}^{(\kappa)}.
  \label{eq:uv-projected-damping-definition}
\end{equation}
From \eqref{eq:uv-frequency-shift}, the leading order solution is then
\begin{equation}
  \gamma_{n,k,\mathrm{proj}}^{(\kappa)}
  =
  \frac{\pi\xi_h}
  {\kappa\left\{
    \left[
      \mathcal L_h^{(\kappa)}
      \bigl(\zeta_n^{(\kappa)}\bigr)
    \right]^2
    +\pi^2\xi_h^2
  \right\}}.
  \label{eq:uv-width}
\end{equation}
Since
\begin{equation}
  \mathcal L_h^{(\kappa)}(\zeta)
  =
  2(1+\xi_h)\log\zeta
  +C_h^{(\kappa)}+o(1),
\end{equation}
where $C_h^{(\kappa)}$ is real, the large-$n$ behavior of the dissipation rate at fixed $k$ is
\begin{equation}
  \gamma_{n,k,\mathrm{proj}}^{(\kappa)}
  \sim
  \frac{\mathcal P(h_R)}
       {\kappa\log^2(n/\kappa)},  
       \label{eq:uv-leading-width}
\end{equation}
where
\begin{equation}
  \mathcal P(h_R)
  =
  \frac{\pi\xi_h}{4(1+\xi_h)^2}
  =
  \frac{\pi h_R^2}{(1+4h_R^2)^2}.
\end{equation}
Thus the projected dissipation rate vanish logarithmically in the radial UV.

\subsection{$c_{\text{UV}}$}

Let $c$ be the central charge of either ambient CFT. We now define a quantity that measures how the conformal interface suppresses the dissipation under the coupling to bath, holding fixed the large radial dimension $\zeta$. Recall $n_\kappa(\zeta)\sim\kappa\zeta$. We then define the central-charge-like quantity by the ratio between the conformal interface at hand and the reference pure AdS$_3$:
\begin{equation}
  \frac{c_{\text{UV}}}{c}
  \equiv
  \lim_{\substack{\zeta\to\infty\\ k\ \mathrm{fixed}}}
  \frac{\gamma_{n_\kappa(\zeta),k,\mathrm{proj}}^{(\kappa)}}
       {\gamma_{n_1(\zeta),k,\mathrm{proj}}^{(1)}}.
  \label{eq:cuv-definition}
\end{equation}
The limit is taken at any fixed real $h_R\neq0$ and with the same
renormalized boundary condition in the numerator and denominator. 

Using \eqref{eq:uv-width}, for BPS-Janus interface its expression is
\begin{align}
  \frac{c_{\text{UV}}}{c}
  &=
  \lim_{\zeta\to\infty}
  \frac{1}{\kappa}
  \frac{
    \left[\mathcal L_h^{(1)}(\zeta)\right]^2
    +\pi^2\xi_h^2}
  {
    \left[\mathcal L_h^{(\kappa)}(\zeta)\right]^2
    +\pi^2\xi_h^2}
  \nonumber\\
  &=
  \frac{1}{\kappa}
  =
  \frac{1}{\cosh\psi\,\cosh\theta}.
  \label{eq:cuv-result}
\end{align}
We can see that in this limit, the coupling-dependent prefactor, temperature of bath, the subtraction
scales, and all finite scheme constants does not appear in the ratio.  The same
answer is obtained for every fixed value of $k$.

% Equivalently, one may compare the same large tower label:
% \begin{equation}
%   \lim_{\substack{n\to\infty\\k\ \mathrm{fixed}}}
%   \frac{\gamma_{n,k,\mathrm{proj}}^{(\kappa)}}
%        {\gamma_{n,k,\mathrm{proj}}^{(1)}}
%   =
%   \frac{1}{\kappa}.
%   \label{eq:cuv-equal-level}
% \end{equation}
% This equal-label prescription differs from \eqref{eq:cuv-definition} at finite
% level, but additive shifts inside the running logarithms disappear in the UV.

For the half-BPS super-Janus family, the known entanglement result gives
$c_{\mathrm{eff}}/c=1/(\cosh\psi\,\cosh\theta)=1/\kappa$
\cite{Gutperle_2016}.  Hence in this limit, the present projected branch
obeys
\begin{equation}
  \frac{c_{\mathrm{UV}}}{c}
  =\frac{1}{\kappa}
  =\frac{c_{\mathrm{eff}}}{c}
  \label{eq:cuv-ceff-equality}
\end{equation}

\section{Conclusion and future directions}
\label{sec:future}

In this work we studied dissipation across a supersymmetric Janus
interface by coupling one asymptotic boundary of the Janus geometry to
a thermal BTZ bath.  The coupling was implemented through a uniform double-trace
deformation on one boundary half-space.  For the top-down massless
scalar, the system operator has dimension two, while the bath field is
treated using the renormalized endpoint alternate quantization with
dimension zero.  The logarithmic Fefferman--Graham expansion on both sides
make it essential to specify the complete renormalized boundary kernel with running scales.

At finite frequency, the Janus and BTZ modes are adapted to different
bases. We specify to the large radial level $n$ with the AdS$_2$ descendant
number $k$ held fixed. 
The off-diagonal elements of the effective bath response matrix are suppressed as
$O(1/(n\log n))$, and we
thereby define and solve the fixed-$k$ scalar projection of the quasinormal modes.

The dissipation rate near the projected normal modes behave as $1/\kappa\log^2(n)$ as we take $n\to\infty$ with fixed $k$.
Although the individual projected damping parameters depend on the
normalization of the double-trace coupling, their ratio to the corresponding
$\kappa=1$ projection does not.  Our universal interface quantity $c_{UV}$ for BPS-Janus is given by
\begin{equation}
  \frac{c_{\text{UV}}}{c}
  =
  \lim_{\substack{\zeta\to\infty\\k\ \mathrm{fixed}}}
  \frac{\gamma_{n_\kappa(\zeta),k,\mathrm{proj}}^{(\kappa)}}
       {\gamma_{n_1(\zeta),k,\mathrm{proj}}^{(1)}}
  =
  \frac{1}{\kappa}
  \label{eq:main-conclusion}
\end{equation}
An immediate consequence of this is that
$\gamma_{n,k,\mathrm{proj}}^{(\kappa)}\to0$ for every fixed $k$ in the
projected scalar problem. Thus, this large-$n$ sector does not exhibit a
finite asymptotic damping gap. Beyond this observation, however, we have not
established how $c_{\text{UV}}$ is related to the Liouvillian gap ratio
$c_{\text{relax}}$ in the free fermion lattice in
\cite{barad2025dissipationmeetsconformalinterface}, and it is left for future work. A natural next step is to numerically calculate the quasinormal modes solved from the full system+bath coupling in \eqref{eq:massless-full-block-matrix} for massless and massive scalars.
As pointed out in the previous section, it would also be interesting to analytically solve the full spectrum of quasinormal modes for this setup, even if only perturbatively in the coupling strength. 

A general recipe for numerically solve the double-trace equations would be restricting to the coupled
half-space, introducing cutoffs, and tracking
the solutions as all cutoffs are increased. However, it is computationally hard to track the behavior of large-frequency solutions as we do not fix the descendant number $k$ to be small.

A second future direction is to change the spatial profile of the system--bath
coupling. A pointlike interaction would most closely parallel the lattice
setup of \cite{barad2025dissipationmeetsconformalinterface}, but there are extra dark modes in addition to the regular quasinormal modes that are subtle to resolve. Smooth nonuniform profiles interpolate between this
localized problem and the uniform half-space coupling considered here.
Determining whether Eq.~\eqref{eq:main-conclusion} survive such
changes would clarify which aspects of $c_{\text{UV}}$ are intrinsic
interface data and which depend on the detailed coupling machanism.

The massless scalar profile is also special. Both the logarithmic $1/\log^2 n$
behavior and the exact cancellation in \eqref{eq:uv-massless-cancellation}
rely on the scalar dimension $(\Delta_J,\Delta_B)=(2,0)$. It would be useful to repeat the
analysis for massive bulk fields, which would enhance the universality of $c_{UV}$.

Finally, it remains to explain directly in field theory why this projected ratio agrees with $c_{\mathrm{eff}}$ in the half-BPS super-Janus family
and to investigate them for less supersymmetric
interfaces. A derivation from defect correlation functions, BOPE data or
spectral sum rules would provide a sharper interpretation of $c_{\text{UV}}$ and establish its connection to $c_\text{relax}$.
More broadly, we would like to understand the relations among
$c_{\text{UV}}$, $c_{\text{relax}}$, $c_{\mathrm{eff}}$, $c_{LR}$, the
boundary entropy $g$, and other dynamical interface observables. It would be
particularly interesting to search for new equalities and bounds such as the ones in \cite{Karch:2024udk}. Connecting closed system quantities like transmission and entanglement entropy with open system quantities like dissipation provides
a dynamical characterization of interfaces.

\section*{Acknowledgements}
We'd like to thank Xueda Wen and Merna Youssef for useful discussions. OpenAI's \texttt{gpt-5.6-sol} model was used to assist with some of the calculations and the writing of Section~4. This work was supported in part by DOE grant DE-SC0022021 and by a grant from the Simons Foundation (Grant 651678, AK).

\bibliographystyle{JHEP}
\bibliography{reference}
\end{document}